\documentclass[twocolumn,preprintnumbers,amsmath,amssymb,superscriptaddress,nofootinbib,longbibliography, prb]{revtex4-2}

\usepackage{graphicx}% Include figure files
\usepackage{bm}% bold math
\usepackage{dsfont}
\usepackage[usenames,dvipsnames]{xcolor}
\usepackage{pstricks}
\usepackage[tight]{subfigure}
\usepackage{verbatim}
\usepackage{units}
\usepackage{multirow}
\usepackage{mathrsfs}
\usepackage{leftidx}
\usepackage{xspace}
\usepackage{diffcoeff}  %\dl x -> dx, \diff[n]{y}{x}[x_0] -> \frac{d^n y}{dx^n}|_{x_0}, \diff*{y}{x} -> \frac{d^n}{dx^n}y, \diffp for partial derivatives
\usepackage{bbm}
\usepackage{amsfonts,amssymb,amsmath}
\usepackage{mathtools}

\usepackage{array} 
\usepackage{tabularx}
\usepackage[normalem]{ulem}
\usepackage{bbold}
\usepackage{pifont}
\usepackage{hyperref}
\hypersetup{colorlinks=true,linktoc=all,linkcolor=blue,breaklinks=true,citecolor=blue,urlcolor=blue}

\usepackage[utf8]{inputenc}
\usepackage[english]{babel}

\definecolor{darkgreen}{RGB}{6, 153, 38}

\addto\captionsenglish{}

\AtBeginDocument{%
    \newwrite\bibnotes
    \def\bibnotesext{Notes.bib}
    \immediate\openout\bibnotes=\jobname\bibnotesext
    \immediate\write\bibnotes{@CONTROL{REVTEX42Control}}
    \immediate\write\bibnotes{@CONTROL{%
    apsrev42Control,author="08",editor="1",pages="1",title="0",year="1"}}
     \if@filesw
     \immediate\write\@auxout{\string\citation{apsrev42Control}}%
    \fi
}%

\newcommand{\beq}{\begin{equation}}
\newcommand{\eeq}{\end{equation}}

\renewcommand*{\H}{\text{H}}

\renewcommand*{\L}{\text{L}}
\newcommand*{\R}{\text{R}}

\DeclarePairedDelimiterX\braket[2]{\langle}{\rangle}{#1\,\delimsize\vert\,\mathopen{}#2}
\DeclarePairedDelimiterX\ketbra[2]{\lvert}{\rvert}{#1\,\delimsize\rangle\mathopen{}\delimsize\langle\,\mathopen{}#2}

\DeclarePairedDelimiterX\Braket[2]{(}{)}{#1\,\delimsize\vert\,\mathopen{}#2}
\DeclarePairedDelimiterX\Ketbra[2]{\lvert}{\rvert}{#1\,\delimsize)\mathopen{}\delimsize(\,\mathopen{}#2}

\newcommand{\kBT}{k_\text{B}T}

\begin{document}

\title{
Mesoscopic routers and single-pole double-throw switches for electronic heat 
}
\author{Jos\'e Balduque}
\affiliation{Departamento de F\'isica Te\'orica de la Materia Condensada, Universidad Aut\'onoma de Madrid, 28049 Madrid, Spain\looseness=-1}
\affiliation{Condensed Matter Physics Center (IFIMAC), Universidad Aut\'onoma de Madrid, 28049 Madrid, Spain\looseness=-1}
\author{Rafael S\'anchez}
\affiliation{Departamento de F\'isica Te\'orica de la Materia Condensada, Universidad Aut\'onoma de Madrid, 28049 Madrid, Spain\looseness=-1}
\affiliation{Condensed Matter Physics Center (IFIMAC), Universidad Aut\'onoma de Madrid, 28049 Madrid, Spain\looseness=-1}
\affiliation{Instituto Nicol\'as Cabrera (INC), Universidad Aut\'onoma de Madrid, 28049 Madrid, Spain\looseness=-1}
\date{\today}

\begin{abstract}
The unavoidable dissipation of heat in electronic nanostructures is a crucial problem, specially when their operation requires low temperatures. It demands finding devices able to control and redirect the excess heat, ideally without perturbing the electrostatic environment. We propose three-terminal junctions working either as thermal routers or as thermal single-pole double-throw switches controlled by a single external knob. Two models are discussed based on resonant tunneling energy filters and different couplings to the heat source: (i) Phase-coherent contact via a scanning tip modulates the relative amount of the two output currents via position-dependent quantum interference; (ii) Coupling via a gate voltage tunable filter selectively switches one of the currents in the presence of dephasing. In the later case, we find that the heat flow using ideal filtering is bounded by fourth the open conductor current. 
%Our results show that the current between two pairs of terminals can be selectively suppressed, leading to highly efficient switching.
\end{abstract}

\maketitle

%%%%%%%%%%%%%%%%%%%%%%%%%%%%%%%%%%%%%%%%%%%%%%%%%%%%%%%%%%%%%%%%%%%%%%%%%%%%%%
%%%%%%%%%%%%%%%%%%%%%%%%%%%%%%%%%%%%%%%%%%%%%%%%%%%%%%%%%%%%%%%%%%%%%%%%%%%%%%
\section{Introduction}
\label{sec:intro}

\begin{figure}[b]
\includegraphics[width=0.7\linewidth]{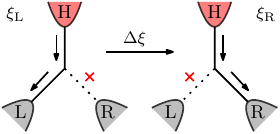}
\caption{\label{fig:SPDT_scheme}
Sketch of an ideal thermal single-probe double-throw switch: tuning an external parameter $\xi$ between two polarities, $\xi_\L$ and $\xi_\R$, simultaneously opens and closes the thermally driven transport between a hot input terminal and two output ones. 
}
\end{figure}

Everybody will agree, without needing to invoke Clausius' statement, on what direction will the heat follow when a cold and a hot body are put in contact. %One would na\"ively expect the answer to be the same even if the cold body is bipartite, formed by two regions at the same temperature. Unfortunately in that case Clausius does not help much, an example being the absorption refrigerator, where heat flows out of the hot and the cold reservoirs to be damped into a third one even in the absence of performed work. 
%The question that we address here is how can one control the flow of heat from a hot source into two different regions which are similarly coupled to it. 
Thermodynamic laws alone are however not enough to predict how this heat will distribute between two regions if the cold body is bipartite.
The question that we address here is how can one use this legal loophole to control the amount of heat that flows into each of these partitions.
The aim is to selectively induce heat to preferentially flow into either of them by tuning an external knob. In that case, the system works as a thermal router. 
%In some limiting cases, the current into one of the regions or the other can be suppressed for at least two values (one for each region) of the knob, so heat flows only into the other one. 
A particularly useful limiting case of this idea is when the knob only allows to switch between two discrete states where heat only flows into one of the regions, being completely suppressed for the other.
Then, the system is a thermal analogue of a single-pole double-throw (SPDT) switch. A schematic representation as a three-terminal junction is sketched in Fig.~\ref{fig:SPDT_scheme}.

To address this question we concentrate on the case of mesoscopic electrical conductors~\cite{ihn_semiconductor_2009}.   
The need to control heat flows in these kinds of systems is demanding: consider for instance the large amount of circuits packed in a chip and dissipating heat while they are operated. This excess heat harms the operation of some surrounding components. Low-power onchip devices have been realized that manage environmental electronic heat by converting it into useful power~\cite{holger,roche:2015,hartmann:2015,jaliel:2019} or cooling~\cite{pekola_colloquium_2021,majidi_heat_2024,balduque_quantum_2026}, rectifying it as thermal diodes~\cite{bipolardiode}, using it as the base signal of a transistor~\cite{thierschmann_thermal_2015}
%li_negative_2006,thierschmann_thermal_2015,sanchez_single_2017,yang_thermal_2019} 
or making it circulate~\cite{granger_observation_2009,nam_thermoelectric_2013}.
%chiraldiode,hwang:2018,acciai:2021,adrian}. 
A thermodynamically consistent description of electronic quantum transport in nanoscale devices is available in terms of scattering theory~\cite{heikkila_book,benenti_thermodynamic_2011}, which allows for a simple analysis of the involved mechanisms. We will restrict to the low temperature regime, where the electron-phonon coupling is suppressed, so heat is transported by electrons.
 
We here propose a three-terminal conductor that redirects (input) heat currents from a hot terminal, H, into two possible output terminals, L and R, switching from one to the other by tuning a single parameter, $\xi$, that represents the knob of the device, cf. Fig.~\ref{fig:SPDT_scheme}. Two models are discussed depending on the nature of the knob: 
%a thermal router is proposed based on a coherent beam splitter where the phase accumulated by electrons along the conductor generates space-dependent and continuously tunable interference patterns in transport; 
a thermal router is proposed based on a coherent beam splitter whose position influences quantum interference processes within the conductor; 
a thermal SPDT switch mechanism is found in an incoherent junction controlled by gate voltage tunable spectral filters.
The ingredients are standard in mesoscopic transport experiments: coherent beam splitters are realized in multiterminal nanowires~\cite{dePicciotto_fourterminal_2001} or scanning probe tips~\cite{binnig1987,voigtlander_miltitip_2018,kovalchuk_imaging_2024}; energy filters appear in resonant tunneling barriers~\cite{chang_resonant_1974} or quantum dots~\cite{josefsson_quantum_2018}.
Interference patterns in coherent one-dimensional conductors have been demonstrated in transport, giving rise to Fabry-Pérot~\cite{liang_fabryPerot_2001,kong_quantum_2001,kretinin_multimode_2010,kotekarPatil_quasibalistic_2017,wang_crossover_2019,ghimire_fabryPerot_2026} and Fano resonances~\cite{kobayashi_tuning_2002,kobayashi_fano_2004} and probed with scanning tunneling microscopes (STM)~\cite{venema_imaging_1999}. 
Their interplay in three-terminal conductors can be modulated by precisely the position of a beam splitter~\cite{balduque_coherent_2024}. 

Single-throw heat switches have been proposed in electronic conductors using magnetic fields in topological Josephson junctions~\cite{sothmann_highly_2017} and in ballistic rings~\cite{haack:2019,haack_nonlinear_2021,adrian}, and realized in superconducting qubits~\cite{ronzani_tunable_2018,maillet_electric_2020}. 
Thermal routers have been realized experimentally using the superconducting phase as a knob in Josephson junctions~\cite{Timossi2018Mar,Guarcello2018Mar} and proposed in the strong magnetic field regime by reflecting chiral edge states in a quantum point contact~\cite{chiraldiode}.
Our proposal uses normal state contacts in the absence of magnetic fields.

Additionally to heat, particle currents flow in response to the increased temperature of terminal H due to nonlocal thermoelectric effects~\cite{balduque_quantum_2026}. In particular, kinetic interference~\cite{hofer:2015,Vannucci2015,genevieve,extrinsic} and energy filtering~\cite{humphrey:02,josefsson_quantum_2018,jordan:2013,jaliel:2019} enable thermoelectric currents. Taking them into account is necessary: on one hand, one can also define a thermoelectric SPDT switch when the three terminals are connected to ground. On the other hand, as heat is actually carried by electrons, the constraints imposed to their flow (whether the terminals are shorted or voltage probes) affects the behaviour of the thermal SPDT. 

The remaining of the manuscript is organized as follows. Section \ref{sec:rout} presents the conditions for a SPDT operation in a coherent electrical conductor. Sections \ref{sec:coherent} and \ref{sec:dephasing} discuss two possible implementations based on the control of the full coherent transmissions via quantum interference, and on spectral filtering with gate voltages, respectively. Section~\ref{sec:conclusions} discusses the conclusions.

\section{Thermally driven current routing}
\label{sec:rout}
%A three terminal system consisting on a hot input terminal and two isothermal output ones provides the minimal configuration to study the routing of particle or heat currents generated from a temperature difference.
%In such scenario, a physical routing mechanism should exist that is able to: i) use the input heat to yield currents of unequal magnitude at the output terminals, and ii) provide some degree of control over this magnitude  by tuning a single external parameter.
%A limiting scenario of this idea is illustrated by the operation principle of an ideal single-probe double-throw, schematized in Fig.~\ref{fig:SPDT_scheme}.
We distinguish between a thermal router and a thermal SPDT switch. The thermal router is a device able to continuously tune the relative amount of heat flowing into the output terminals, $o=\L,\R$. Ideally it ranges between two values of the control parameter, $\xi=\xi_o$, where current flows into only one terminal, $o$, with the other one being opaque to heat.
A SPDT device is expected to be able to switch between the two states $\xi_\L$ and $\xi_\R$ in a discrete way: even if $\xi$ is necessarily a continuous parameter, there are no states for which heat flows into both output terminals simultaneously.
%
%A proper SPDT switch operation will hence be characterized by the degree of suppression of the switched off current. Also by the quality of the switching process. If the parameter $\xi$ is continuous, heat will unavoidably leak into both output terminals. To mitigate this, either the two values $\xi=\xi_\L$ and $\xi=\xi_\R$ for which heat only flows into L and R are close to each other, or the switching is done very fast. 
While we are here not considering the dynamical process of switching, we will get some notion of this effect from how the currents change as $\xi$ is continuously tuned.    
%Real systems will nevertheless present a transient regime between these two fully directional states where a finite current flow at both output terminals simultaneously, which should be properly characterized, and can provide the system with more complex commutation operations, as we will discuss below.
%We will also distinguish between strict routing, where the current always flows from the input terminal to the output ones, from more general scenarios where the currents do not have a definite direction. \textcolor{red}{(?)}

In multi-terminal mesoscopic systems where correlations due to electron-electron interactions can be disregarded, transport is described by the scattering matrix ${\cal S}_{jj'}(E)$ informing on the transmission amplitudes for electrons emitted by terminal $j'$ and absorbed by terminal $j$~\cite{landauer_spatial_1957,buttiker_four_1986}. 
These amplitudes are sensible to the phases accumulated by the electrons along their trajectories between terminals, which can result in interference patterns in the transmission probabilities ${\cal T}_{jj'}(E)=|{\cal S}_{jj'}(E)|^2$.  
We are already assuming here for simplicity that electrons are injected via single-channel leads.  
This will be the case in the rest of the manuscript except when explicitly mentioned.
The generated particle, $I_j=\int dE{\cal I}_j(E)$ and heat, $J_j=\int dE(E-\mu_j){\cal I}_j(E)$, currents flowing out of terminal $j$ are calculated with the transmission probabilities via the spectral particle current~\cite{sivan:1986}
%well described by the Landauer-B\"uttiker scattering theory~\cite{moskalets-book}, which considers completely coherent transport between terminals according to a scattering matrix $\mathcal{S}_{jj'}$.
%For a single transport channel per lead they read $I_j=\int dE{\cal I}_j(E)$ and $J_j=\int dE(E-\mu_j){\cal I}_j(E)$ in terms of the spectral current
%\begin{align}
%\label{eq:Ipart}
%I_j&=\frac{2}{h}\sum_{j'}\int{dE} {\cal T}_{j'j}(E)[f_j(E)-f_{j'}(E)]\\
%\label{eq:Jheat}
%J_j&=\frac{2}{h}\sum_{j'}\int{dE} (E-\mu_j){\cal T}_{j'j}(E)[f_j(E)-f_{j'}(E)],
%\end{align}
\begin{align}
\label{eq:IE}
{\cal I}_j(E)=\frac{2}{h}\sum_{j'}{\cal T}_{jj'}(E)[f_j(E)-f_{j'}(E)],
\end{align}
where 
%${\cal T}_{j'j}(E)=|{\cal S}_{j'j}(E)|^2$ is the transmission probability from $j$ to $j'$ and 
$f_j(E)=\{1+\exp[(E{-}\mu_j)/\kBT_j]\}^{-1}$, is the Fermi distribution function of terminal $j$ having an electrochemical potential $\mu_j$ and a temperature $T_j$. 
The factor 2 in Eq.~\eqref{eq:IE} accounts for spin degeneracy in the absence of a magnetic field, in which case  we also have ${\cal T}_{j'j}(E)={\cal T}_{jj'}(E)$~\cite{moskalets-book}. We will consider that the temperature is homogeneous in the output terminals, $T_\L=T_\R=T$, with terminal H being hot, $T_\H>T$.
The same processes will however apply in the reverse configuration where $T_\H<T$, so one can select which terminal (L or R) is used to inject heat into H.

%The coupling of particle and heat currents introduces an additional degree of freedom. 
The fact that heat is carried by electrons introduces an additional degree of freedom. 
With their temperatures fixed, the particle flow will depend on how the different terminals are wired among them and to the ground. For instance, due to the thermoelectric effect, a terminal $j$ being a voltage probe will develop an electrochemical potential $\mu_j$ such that the average particle current vanishes~\cite{buttiker:1986}, $I_j=0$, though a finite heat current is allowed via fluctuations.
%On top of heat, the temperature difference in the input generates an electrical response due to the thermoelectric effect~\cite{balduque_quantum_2026}.
We account for this by considering three operation modes for the device according to the boundary conditions imposed to the input and output terminals, which represent three physically relevant scenarios: (i) {\it short circuit}, where all chemical potentials are fixed and equal, $\mu_j=\mu$, so finite particle currents flow through the three terminals, (ii) {\it voltage-probe heat source}, where the hot input terminal is galvanically isolated from the ground, with a floating electrochemical potential $\mu_{\rm H}$ that maintains $I_{\rm H}(\mu_{\rm H})=0$, but $I_{\rm L}=-I_{\rm R}\neq0$, in general~\cite{balduque_coherent_2024}, and (iii) {\it open-circuit} or {\it all-thermal} configuration, where all terminals are voltage probes, so none of them injects particles, i.e., $I_{\rm L}=I_{\rm R}=I_{\rm H}=0$. The heat currents will be different in each case.  For concreteness, we will mostly focus on the short circuit configuration and compare the three cases for the coherent injection model in Sec.~\ref{sec:coherent}.

On top of this, one can also think on the device working to route the particle current by means of a temperature increase in H. In this case one has a thermoelectric router. We explore this possibility in appendix~\ref{app:particleSPDT}.
Note that for the particle currents, the router only works when all electrochemical potentials are fixed, in this case corresponding to the short circuit configuration. 

In what follows, we will explore two different routing mechanisms that rely on externally modifying either the transmission probabilities, or the electron distribution at some region of the conductor.
In the first case, we use the tip of an STM to influence quantum interference processes within a fully coherent conductor.
In the second, we use a gate-tunable energy filter to selectively match the local spectral properties of the different output contacts. 

\section{Routing heat via interference
%Coherent model
}
\label{sec:coherent}

%%%%%%%%%%%%%%%%%%%%%%%%%%%%%%
\begin{figure}[t]
\includegraphics[width=\linewidth]{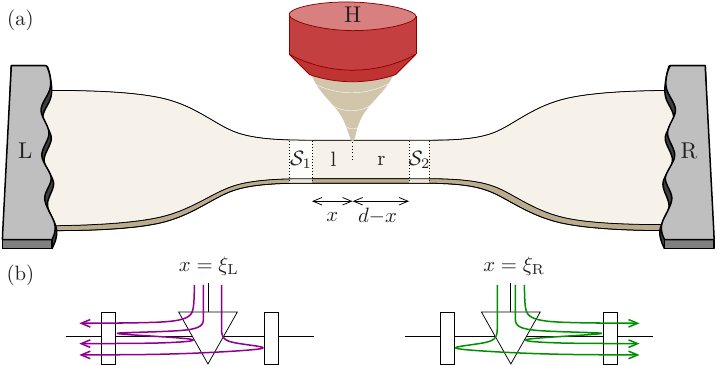}
\caption{\label{fig:scheme}
Sketch of a tunable three terminal conductor. (a) A scattering region connects two terminals, L and R, and is coupled to a third terminal, H, via a scanning tip. The scattering region is composed of two resonant tunneling barriers described by scattering matrices ${\cal S}_\alpha$ and separated by a ballistic channel of length $d$. 
The tip position, $x$, divides the separation between barriers in two branches, l and r.
(b) For particular positions of the tip ($x=\xi_\L$ and $x=\xi_\R$) heat flows into only one output terminal (L or R, respectively) due to the destructive interference of trajectories ending in the other one.}
\end{figure}
%%%%%%%%%%%%%%%%%%%%%%%%%%%%%%
We first consider the routing properties of a fully coherent system where the switching parameter modulates the interference pattern of the transport probabilities. For concreteness we consider the system sketched in Fig.~\ref{fig:scheme}, where heat is injected from terminal $\H$ into the conductor (e.g., a quasi-one-dimensional quantum wire) via a ballistic beam splitter~\cite{buttiker:1984,buttiker:1989} of tunable position. 
Such junction can be physically implemented using the tip of an scanning tunneling microscope (STM)~\cite{pugnetti_electron_2009,shastry_scanning_2020,genevieve,extrinsic,cioni_high_2025,fonck_quantum_2025}. 
%It consists of two output terminals, $j=\rm L,R$, having electrochemical potentials $\mu_j$ and being at the same temperature, $T$, and an input terminal, H, at temperature $T_{\rm H}=T+\Delta T_{\rm H}$ and electrochemical potential $\mu_{\rm H}$, coupled to the conductor between them. 
The conductor contains two scattering centers, labeled  $\alpha=1,2$, separated by a distance $d$ scanned by the tip. %The internal reflections in the separation of 1 and 2 defines the interference of trajectories with different accumulated kinetic phases.   
We assume the scattering centers to be resonant tunneling barriers of resonance energy $\varepsilon_\alpha$, and broadening $\Gamma_\alpha$, with their transmission amplitudes $\tau_\alpha(E)$ given by Breit-Wigner resonances~\cite{buttiker:1988}. They contribute to transport as Lorentzian peaks of the form
\beq
|\tau_\alpha(E)|^2=\Gamma_\alpha^2/[(E-\varepsilon_\alpha)^2+\Gamma_\alpha^2].
\label{eq:resonance}
\eeq
For simplicity we will in the following consider the broadening of all resonances to be the same, $\Gamma_\alpha=\Gamma$.
%We consider a minimal model of the conductor as a single one-dimensional  elastic channel embedded with two resonant tunneling scatterers, $\alpha=1,2$, of resonance energy $\varepsilon_\alpha$, and equal broadening $\Gamma$. 

We model the tip-conductor junction to symmetrically inject waves from H into the left and right branches in the wire. This way, its scattering matrix is (up to phases) determined by a single parameter, $\epsilon\in[0,1/2]$, which defines the strength of the tip-conductor coupling~\cite{buttiker:1984}.
This setup has the property of allowing the heat source to be disconnected ($\epsilon=0$) by simply separating the tip, and has a dual operation as an efficient and tunable heat to work converter~\cite{balduque_coherent_2024}. We refer the reader to this reference for technical details of the scattering properties.

The position of the tip with respect to the two resonant tunneling scatterers defines two internal branches, l and r, of length $x$ and $d-x$, where electrons can undergo multiple phase-coherent reflections while accumulating kinetic phases 
$\chi_{\rm l}(E)=2k(E)x$ and $\chi_{\rm r}(E)=2k(E)(d-x)$, respectively.
Here, $k(E)=\sqrt{2m^*(E-U_0)}/\hbar$ is the electron wave-number, with $U_0=0$ the energy of the lowest subband of the one-dimensional conductor and $m^*$ the effective electron mass~\cite{ihn_semiconductor_2009}.
Tuning the position of the tip changes the relative lengths of branches l and r and therefore the resulting interference processes, as sketched in Fig.~\ref{fig:scheme}(b). 
Hence, we can use this position as a knob parameter, $\xi=x$.
In the following we will express distances in terms of the thermal length $l_0=\hbar/\sqrt{8m^*\kBT}$.

It is important to emphasize that the interface between branches l and r established by the position of the tip has an associated reflection amplitude, $\eta_-=[1-\sqrt{1-2\epsilon}]/2$, for electrons propagating along the wire.
Each branch separately can hence be viewed as an electronic Fabry-P\'erot interferometer formed by the reflections at one of the scattering regions and at the tip interface. They result in constructive interference with resonant energies
\begin{equation}
\label{eq:resonant_energies}
    \varepsilon_{{\rm l}n}=\frac{(2n-1)^2\pi^2}{(x/l_0)^2}, \quad \varepsilon_{{\rm r}n}=\frac{(2n-1)^2\pi^2}{[(d-x)/l_0]^2}, 
\end{equation}
where $n$ is an integer, consequent with the resonant condition $\cos[\chi_{\rm l(r)}(E)]=-1$~\cite{Datta1995}. We will refer to these as {\it partial} resonances. 

The transmission amplitude for electrons going between branches l and r across the tip interface is $\eta_+=[1+\sqrt{1-2\epsilon}]/2$. Trajectories for electrons injected from H and absorbed by terminal $o$ = L(R) going through branch $\sigma$ = l(r) may also visit the other branch, $\sigma'$ = r(l) and be reflected at region $\alpha'=2(1)$. In cases where $|\tau_{\alpha'}(\varepsilon_{\sigma'n})|^2\ll1$, this branch acts as a localized state coupled in parallel to the main transport channel, a configuration leading to Fano resonances~\cite{fano_spettro_1935,fano_effects_1961} in transport~\cite{tekman_fano_1993}.

%%%%%%%%%%%%%%%%%%%%%%%%%%%%%%%%%%
\begin{figure*}[t]
\centering
\includegraphics[width=0.7\linewidth]{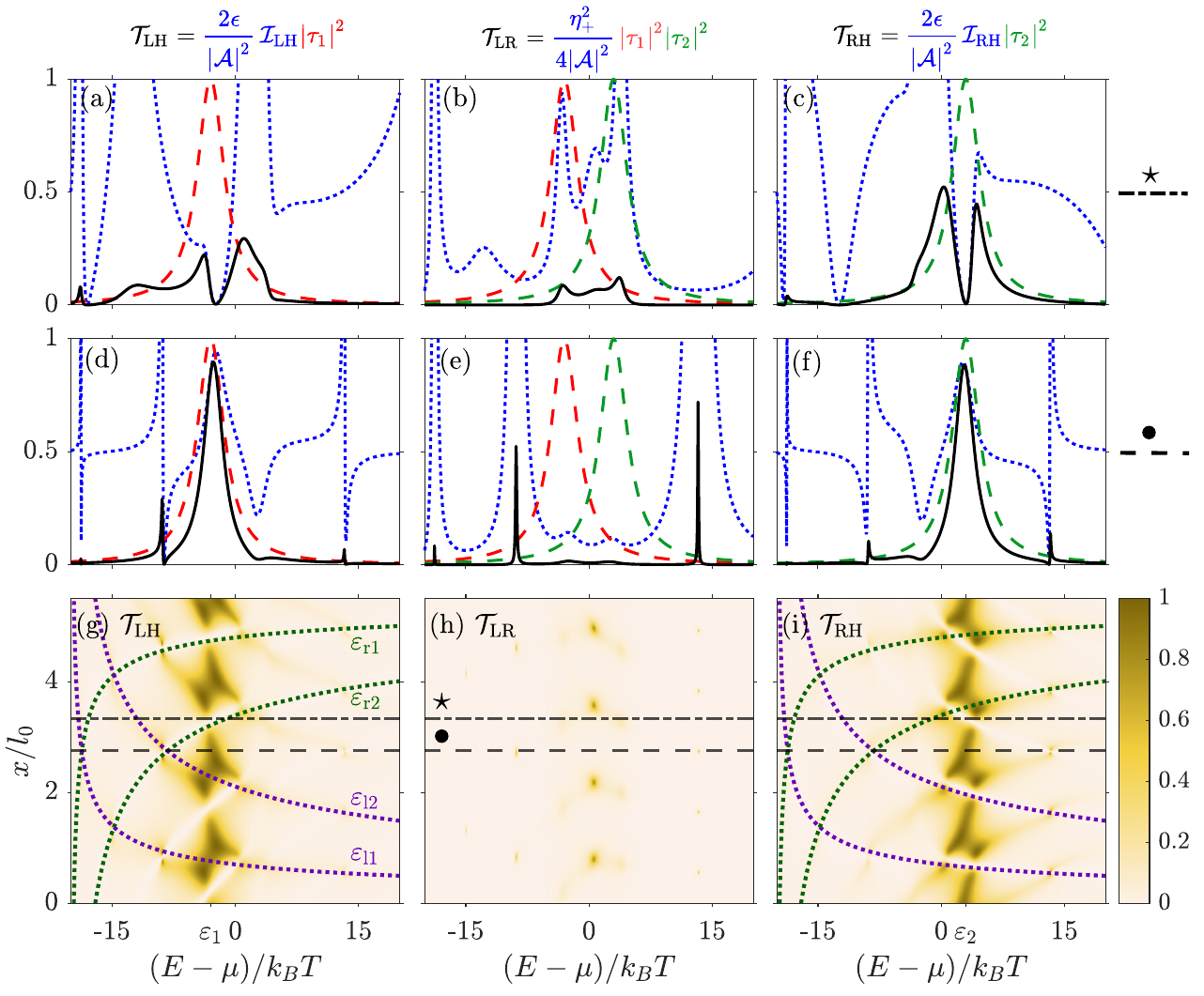}
\caption{\label{fig:trans}  Transmission probabilities between (a),(d) terminals L and H, (b),(e) terminals L and R and (c),(f) terminals R and H as a function of the energy of the scattered electrons for tip positions marked in panels (g)-(i) with $\star$ and $\bullet$ symbols. The total transmission (black full line) can be decomposed as suggested in Eqs. \eqref{eq:transmissionslh} and \eqref{eq:transmissionslr} into the contribution of the resonant tunneling barriers, $|\tau_1|^2$ (red-dashed lines) and $|\tau_2|^2$ (green-dashed lines), and the contribution due to the multiple internally reflected trajectories between scattering regions and the tip (blue-dotted lines). (g)-(i) Color maps of the previous as a function of also the tip position. Dotted purple and green lines mark the partial-resonance conditions ($n=1$ and $n=2$) for l and r internal branches, respectively, as given by Eq.~\eqref{eq:resonant_energies}. Dashed-dotted and dashed lines indicate the tip position used in panels (a)-(c) and (d)-(f) respectively. Parameters: $d=5.5l_0, \; \epsilon=0.5,  \; (\varepsilon_1-\mu)=-3\kBT, \; (\varepsilon_2-\mu)=3\kBT, \; \Gamma = 2\kBT, \; \mu=20\kBT$.}
\end{figure*}
%%%%%%%%%%%%%%%%%%%%%%%%%%%%%%%%%%

The total transmission probabilities are calculated by combining the different scattering regions, reading~\cite{balduque_coherent_2024}
\begin{equation}
\label{eq:transmissionslh}
    \mathcal{T}_{o \rm H}(E) =2\epsilon|\mathcal{A}|^{-2}|{\tau}_{\alpha_o}|^2\mathcal{I}_{o \rm H}, 
\end{equation}
for trajectories between terminals $o=\L,\R$ and terminal H, with $\alpha_{\rm L(R)}$=1(2), and
\begin{equation}
\label{eq:transmissionslr}
    \mathcal{T}_{\rm LR}(E) = \frac{\eta_{+}^2}{4|\mathcal{A}|^2}|{\tau}_1|^2|{\tau}_2|^2,
\end{equation}
for the elastic transport between output terminals. 
%In these expressions, $\tau_\alpha$ are the transmission amplitudes at scatterer $\alpha$, given by Breit-Wigner resonances~\cite{buttiker:1988} resulting in
%\beq
%|\tau_\alpha(E)|^2=\Gamma_\alpha^2/[(E-\varepsilon_\alpha)^2+\Gamma_\alpha^2],
%\label{eq:resonance}
%\eeq
%while
In these expressions,
\begin{equation}
\label{eq:MechanicalPhE_A}
    \mathcal{A}=  1+\frac{\eta_{-}}{2}\left(r_1e^{i\chi_{\rm l}}{+}r_2e^{i\chi_{\rm r}}\right) - \sqrt{1{-}2\epsilon} \; r_1 r_2 e^{i\chi_d}
\end{equation}
and
\begin{equation}
\label{eq:MechanicalPhE_Interference_pattern}
\mathcal{I}_{j \rm H} =  1-\frac{|\tau_{\beta_{j}}|^2}{2} + {\rm Re}[r_{\beta_{j}} e^{i\chi_j}]
\end{equation}
account for the (energy dependent) quantum interference processes, with $r_\alpha=1+\tau_\alpha$ the reflection amplitudes at scattering center $\alpha$, $\chi_d=\chi_{\rm l}+\chi_{\rm r}$, $\chi_{\rm L(R)}=\chi_{\rm r}(\chi_{\rm l})$ and $\beta_{\rm L(R)}$=2(1). As the scattering matrix is unitary, the reflection probabilities can be calculated from the transmissions: $\mathcal{T}_{jj}=1-\sum_{j'\neq j}\mathcal{T}_{jj'}$. 

\subsection{Unraveling the interference contribution}
We illustrate the properties of the transmission probabilities and the interference pattern in Fig.~\ref{fig:trans}.
Figures~\ref{fig:trans}(g)-(i) show a colormap of the three transmission probabilities as functions of the energy of the scattered electrons and the tip position.
The elastic transmission between L and R is strongly suppressed throughout the map for the chosen parameters. As expected, the transmission of electrons from the tip to the conductor terminals L and R peaks around the resonance energies of the scatterers, $\varepsilon_1$ and $\varepsilon_2$, respectively. 
Additionally, we observe features coinciding with the partial resonance conditions of Eq.~\eqref{eq:resonant_energies}: ${\cal T}_{\rm LH}$ exhibits narrow peaks at $\varepsilon_{{\rm l}n}$ as well as suppressions at $\varepsilon_{{\rm r}n}$, and viceversa for ${\cal T}_{\rm RH}$.
%Furthermore, both $\mathcal{T}_{\rm LH}$ and $\mathcal{T}_{\rm RH}$ display vanishing spots at the resonant energies $\varepsilon_{{\rm l}n}$ and $\varepsilon_{{\rm r}n}$, respectively. 
These conditions are highlighted in Fig.~\ref{fig:trans}(g) and Fig.~\ref{fig:trans}(i) as dotted purple and green lines.
We attribute the suppressions to the contribution of Fano-like interference: when the partial resonance condition is fulfilled in branch l(r), the scattering wave function becomes strongly localized there, suppressing the transmission between the tip and terminal R(L).

This is more clearly visible in Figs.~\ref{fig:trans}(a)-(f), where we plot the energy dependence of the transmissions for fixed tip positions [indicated with dashed and dashed-dotted lines in panels \ref{fig:trans}(g) and \ref{fig:trans}(i)]. 
The total transmissions are shown as solid black lines. To better understand the underlying mechanisms, we factorize the transmission expressions of Eqs.~\eqref{eq:transmissionslh} and \eqref{eq:transmissionslr}: we plot the contributions arising from the resonant scatterers $|\tau_\alpha|^2$, giving Lorentzian line shapes around the energies $\varepsilon_1$ (red-dashed line) and $\varepsilon_2$ (green-dashed line), as well as the interference pattern $2\epsilon{\cal I}_{o\H}|{\cal A}|^{-2}$ or $\eta_+^2|{\cal A}|^{-2}/4$ for the corresponding cases (blue-dotted lines).

Let us focus first on Fig~\ref{fig:trans}(a): the overall transmission ${\cal T}_{\L\H}(E)$ is confined to the region dominated by the resonance $\varepsilon_1$, which acts as a bottleneck for transport. The transmission is however modulated by the interference pattern, which shows sharp resonances at the partial-resonance energies of branch l [when the dotted-purple lines  cross the dashed-dotted line in Fig.~\ref{fig:trans}(g)]. 
We attribute these features to Fabry-P\'erot-like constructive interference. 
In contrast, at the partial-resonance energies of branch r [when the dotted-green lines cross the dashed-dotted line in Fig.~\ref{fig:trans}(g)], the interference pattern becomes destructive as expected, completely canceling the total transmission.
The opposite situation is observed for $\mathcal{T}_{\rm RH}$ in Fig.~\ref{fig:trans}(c): sharp resonances appear at $\varepsilon_{\rm r}$, while $E=\varepsilon_{\rm l}$ results in destructive interference. 
In cases when both resonant conditions occur simultaneously, e.g., for the case with $d=x/2$ [marked by a black dashed line in Figs~\ref{fig:trans}(g) and \ref{fig:trans}(i)] the interference pattern is dominated by very clear Fano-like resonances as shown in Figs~\ref{fig:trans}(d) and \ref{fig:trans}(f). 
Consistently, the Fano line-shapes occur in energy regions where one of the barrier transmissions $|\tau_\alpha|^2$ is suppressed.
Lastly, from Figs.~\ref{fig:trans}(b) and \ref{fig:trans}(e) we can attribute the small values of $\mathcal{T_{\rm LR}}$ to the small overlap between the two Lorentzian line-shapes, which is also affected by the internal reflection processes.

\begin{figure}[t]
\centering
\includegraphics[width=\linewidth]{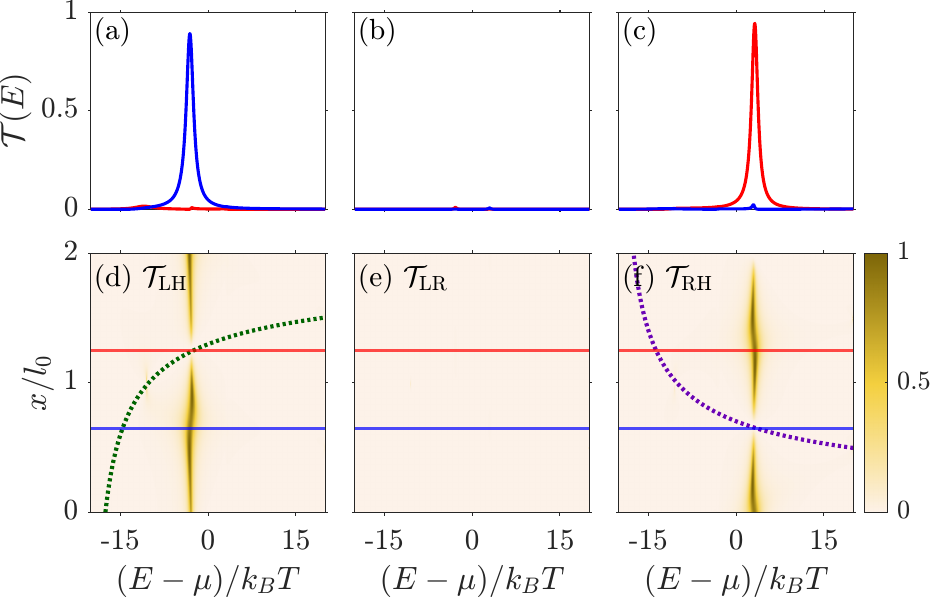}
\caption{\label{fig:MechanicalPhE_valve_trans} Transmission probabilities between (a) terminals L and H, (b) terminals L and R and (c) terminals R and H as a function of the energy of the scattered electrons, and for two different tip positions indicated by red and blue horizontal lines in panels (d)-(f). (d)-(f) Color maps of the previous as a function of also the tip position. Dotted purple and green lines mark the partial resonance conditions ($n=1$) for internal branches l and r, respectively, as given by Eqs.~\eqref{eq:fano_positionL} and \eqref{eq:fano_positionR}. Parameters: $d=2l_0, \; \epsilon=0.5,  \; \varepsilon_1-\mu=-3\kBT, \; \varepsilon_2-\mu=3\kBT, \; \Gamma = 0.5\kBT, \; \mu=20\kBT$.}
\end{figure}

\subsection{Position-dependent modulation of transport}
The position of the tip can then be used to change the transmission shape at will by modifying the interference term, in particular suppressing transport between the output terminals, as required for a current router. 
We show a particular example for $d=2l_0$ in Fig.~\ref{fig:MechanicalPhE_valve_trans}. 
By placing the tip at the positions that fulfill $\varepsilon_{\rm{l}}(x)=\varepsilon_{2}$ and $\varepsilon_{\rm{r}}(x)=\varepsilon_{1}$, indicated by blue and red horizontal lines in Figs.~\ref{fig:MechanicalPhE_valve_trans}(d) and \ref{fig:MechanicalPhE_valve_trans}(f), we can completely block the transport in one of the output terminals, and force it to occur between H and the other remaining one, see Figs.~\ref{fig:MechanicalPhE_valve_trans}(a)-(c).
Hence, we can switch between two configurations where the transport is fully routed from the heat source to one of the conductor terminals by simply sweeping the tip position.
We label this switching positions as
\begin{equation}
    \label{eq:fano_positionL}
    x_{{\rm L}n} =\frac{(2n-1)\pi}{\sqrt{\varepsilon_{2}}}l_0
\end{equation}
for transport being blocked at terminal L and
\begin{equation}
    \label{eq:fano_positionR}
    x_{{\rm R}n} =d-\frac{(2n-1)\pi}{\sqrt{\varepsilon_{1}}}l_0
\end{equation}
for transport being blocked at terminal R.

Remarkably, the transmission ${\cal T}_{\L\R}$ vanishes for all energies and tip positions, see Fig.~\ref{fig:MechanicalPhE_valve_trans}(e): the two wire terminals become effectively decoupled, so transport only occurs via either of them and the input terminal, H. This has the additional advantage of helping to reduce the output current fluctuations.

\begin{figure}[t]
    \centering
    \includegraphics[width=.9\linewidth]{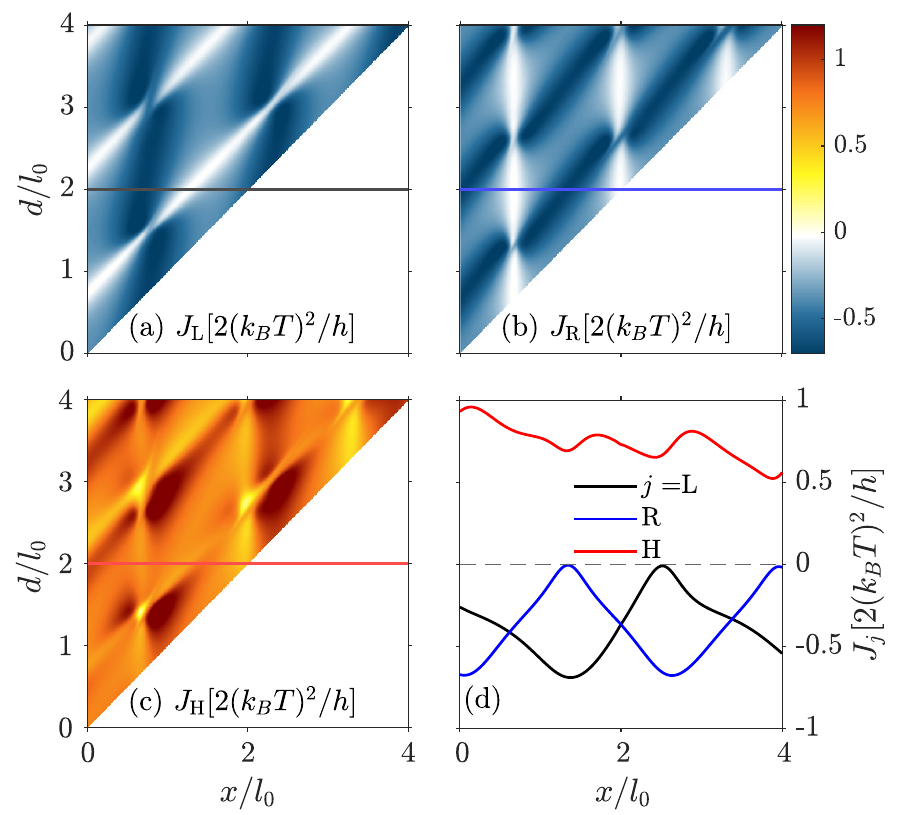}
   \caption{\label{fig:currents_d_vs_x} (a)-(c) Heat currents in reservoirs $j=$L,R,H, when the temperature of reservoir H is increased by $\Delta T/T=1$, as a function of the separation between scatterers $d$ and the tip position $x$. Panel (d) shows cuts of the previous for fixed distance $d=2l_0$, indicated with horizontal lines. Parameters: $\varepsilon_1-\mu=-3\kBT$, $\varepsilon_2-\mu=3\kBT$, $\epsilon=0.5$, $\Gamma=0.5\kBT$ and $\mu=20\kBT$. }
\end{figure}

\begin{figure*}[t]
    \includegraphics[width=.8\linewidth]{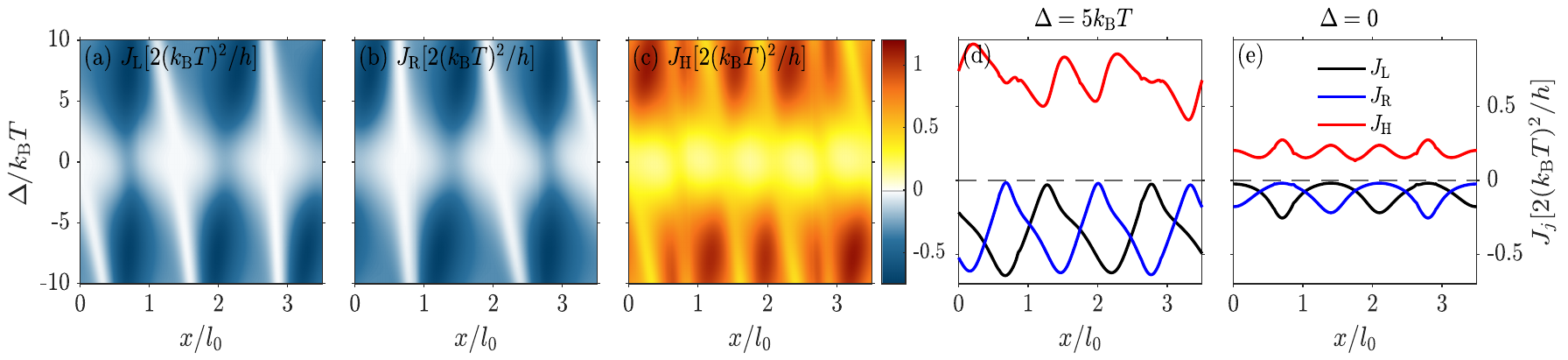}
   \caption{\label{figcurrent_DeltaE_vs_x} (a)-(c) Heat currents in reservoirs $j=$L,R,H, when the temperature of reservoir H is increased by $\Delta T/T=1$, as a function of the tip position $x$, and the energy gain $\Delta$, with $\varepsilon_1+\varepsilon_2=2\mu$. Panels (d) and (e) show cuts of the previous for fixed energy gains $\Delta=5\kBT$ and $\Delta=0$, respectively. Parameters: $d/l_0=3.5$, $\epsilon=0.5$, $\Gamma=0.5\kBT$ and $\mu=20\kBT$. }
\end{figure*}

These features are directly transferred to the particle and heat currents generated when heating the input terminal. They are plotted in Fig.~\ref{fig:currents_d_vs_x} as functions of the distance between scatterers and the tip position. The energy of the quantum dot resonances is here tuned symmetrically with respect to the electrochemical potential, with their energy splitting being $\Delta\equiv\varepsilon_2-\varepsilon_1$.
As expected, the destructive interference manifests as $x$-periodic nodes for the heat output currents, appearing around the positions determined by Eqs.~\eqref{eq:fano_positionL} and \eqref{eq:fano_positionR}. Quite conveniently, these nodes extend over large ranges of the systems size $d$, making the effect robust on the design of the system: only for particular configurations the switch effect is absent. 
The heat current is always injected by the tip and divided between both output terminals, see Figs.~\ref{fig:currents_d_vs_x}(a)-\ref{fig:currents_d_vs_x}(d). Moving the position of the tip one can chose points where whether the heat injected from H flows into L or into R, with the amount of transferred heat being the same in both cases, see Fig.~\ref{fig:currents_d_vs_x}(d).
In the way between these positions, the ratio of the two currents can be continuously and smoothly controlled.
%On the contrary, the energy difference $\Delta$ determines a preferred direction for the thermoelectric current, so that, when finite, particles always leave the left reservoir and enter the right one~\cite{balduque_quantum_2026}, as shown in Figs.~\ref{fig:currents_d_vs_x}(a)-\ref{fig:currents_d_vs_x}(d).
%\textcolor{red}{Hence, this configuration allows to switch between an scenario where current is injected from L and solely transferred to H, to another where current is injected from H and solely transferred to R, enabling the actuated circulation of particles. (esto era para el caso de las partículas)}

In an actual experiment, the distance between scatterers is fixed (e.g., controlled by the vapour deposition times in the wire growth~\cite{ihn_semiconductor_2009}), while the resonant energies are easily controllable with voltages applied to plunger gates. 
To take this limitation into account we plot the currents as functions of $\Delta$ and $x$ for a fixed distance $d=3.5l_0$ in Fig.~\ref{figcurrent_DeltaE_vs_x}. We furthermore consider the barrier resonance energies to be anti-symmetrically situated with respect to the electrochemical potential, i.e., $\varepsilon_1+\varepsilon_2=2\mu$. We again observe the strong suppression of the output currents at different positions of the tip, see Figs.~\ref{figcurrent_DeltaE_vs_x}(a)-\ref{figcurrent_DeltaE_vs_x}(d). 
Compared to Fig,~\ref{fig:currents_d_vs_x}, the current suppression is no longer perfectly periodic with $x$, since the destructive interference conditions~\eqref{eq:fano_positionL} and~\eqref{eq:fano_positionR} also depend on $\Delta$. 
%\sout{The later also dictates the sign of the thermoelectric current, see Fig.~\ref{figcurrent_DeltaE_vs_x}(a)-\ref{figcurrent_DeltaE_vs_x}(c).}

As a particularly interesting configuration, we include in Fig.~\ref{figcurrent_DeltaE_vs_x}(e) the cuts for fixed $\Delta = 0$, where the two quantum dot resonances completely overlap.
The contribution of these resonances to the transmission coefficients, $|\tau_1(E)|^2$  and $|\tau_2(E)|^2$, corresponds to two peaks centered at the equilibrium electrochemical potential, $\mu$~\cite{symres}. Electrons propagating at these energies carry small amounts of heat: the thermal response of the two quantum dots is limited by their broadening, $\Gamma$.
This explains the suppression of the heat currents for most of the tip positions, with small peaks where current alternately flows into L or R, see Fig.~\ref{figcurrent_DeltaE_vs_x}(e). 
%\textcolor{red}{Hence, the finite thermoelectric currents can only be attributed to the energy dependence introduced by the interference phenomena, a feature previously explored in Refs.~\cite{extrinsic,balduque_coherent_2024}}. 
%They are originated by the tip position forming two Fabry–Pérot interferometers (in both branches l and r) of tunable length. 
The different peaks correspond to the appearance of a resonance either in branch l or r near the equilibrium electrochemical potential which hybridizes with the barrier resonances and consequently increases the flow of particles between H and terminals L or R, respectively. 
%These resonances are typically symmetric, so the electron-hole symmetry is preserved when they align with $\mu$, therefore providing a heat current peak and a vanishing particle current (not shown). 
The conditions for the appearance of Fano interference (a pseudo-localized state in a closed branch is coupled in parallel to a transport channel along the other branch) are not met in this case: for energies contributing to transport, both l and r branches are resonantly connected to the leads via barriers 1 and 2. For other energies (far from $\mu$), both branches are closed and there is no transport. 
%The positions of the peaks coincide with the vanishing of the corresponding particle currents (not shown), as expected for the contribution of single resonances.
%This again results in a switching behaviour that now (rather than on destructive interference) relies on constructive interference effects.
The switching effect is in this case weaker and results from constructive (rather than Fano-induced destructive) interference.

\subsection{Characterization of performance}
\label{sec:performance}
\begin{figure}[t]
%    \centering
    \includegraphics[width=\linewidth]{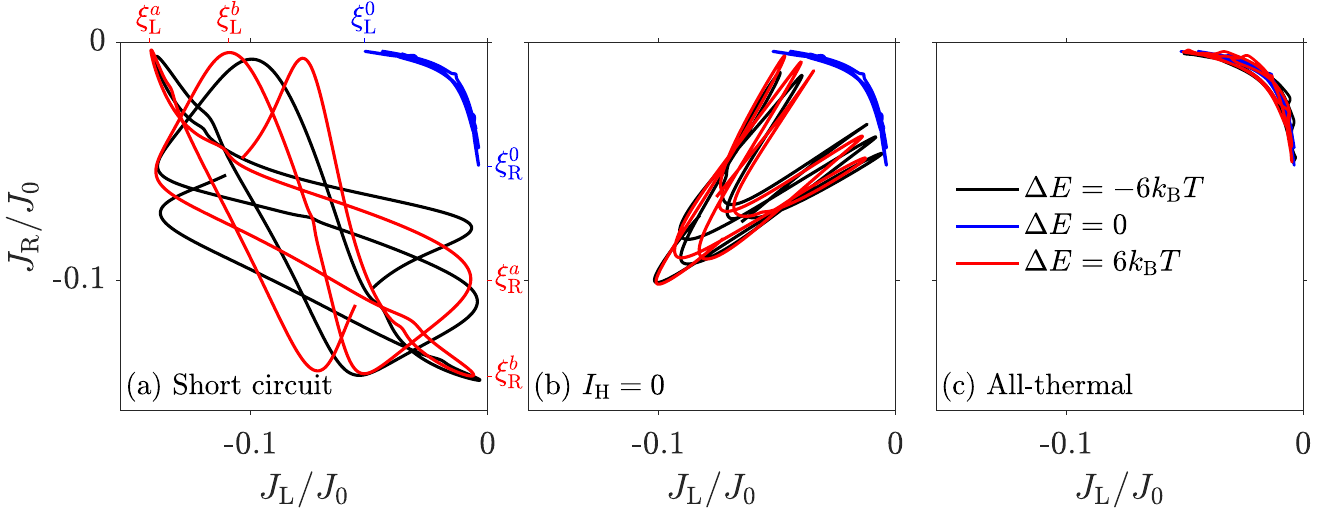}
   \caption{\label{fig:heat_lazos} 
   Heat currents as the position of the tip position, $x$ is tuned, with fixed $d/l_0=3.5$ and different $\Delta$ and $\varepsilon_1+\varepsilon_2=2\mu$ for (a) \textit{short circuit}, (b) \textit{voltage-probe heat source} and (c) \textit{all-thermal} configurations. Other parameters as in Fig.~\ref{figcurrent_DeltaE_vs_x}.
   }
\end{figure}

The desired properties of the router are that the knob covers all the ratios of the output currents when tuned between the two switch points, $\xi_\L$ and $\xi_\R$, doing so in a smooth and predictable way and maintaining the input $J_\H$. Also at the $\xi_o$ points, the suppression of one of the currents should be strong and the response the system, symmetric, i.e., $J_\L$ at $\xi_\L$ should be equal to $J_\R$ in $\xi_R$. Additionally the junction would preferably have a low thermal resistance, allowing for currents as large as possible, ideally as large as the current flowing by a (spin degenerate) open channel, 
\beq
J_0=\pi^2 k_{\rm B}^2(T_\H^2-T^2)/3h. 
\eeq
%\textcolor{red}{Conversely, for the thermoelectric case, a large Seebeck coefficient}. 
Finally, one can also request that the switching is not restricted to specific points but tolerates inaccuracies in the switching parameter (in this case, the position of the tip). 

To have a more complete picture of the characterization of the switch, we show in Fig.~\ref{fig:heat_lazos}(a) a parametric plot for the output heat currents as the tip position is swept from $x=0$ to $x=d$. 
We can distinguish two radically different behaviours depending on wether $\Delta$ is finite or not, which evidences the different underlying switching mechanisms discussed before.
For $\Delta=0$ the response stays close to the axes where one of the currents vanishes, resulting in a high tolerance. 
Furthermore, the input current presents oscillations around a constant value, see also Fig.~\ref{figcurrent_DeltaE_vs_x}(e). 
Hence, the process of switching between the two switch points, labeled $\xi_\L^0$ and $\xi_\R^0$, can be done swiftly without involving large heat currents leaking into the two junctions, a behaviour closer to that of a thermal SPDT switch.
However the allowed output currents are small, and the amount of accessible ratios limited.

In contrast, for $\Delta\neq 0$ the current is switched off in various different points for each terminal. This is an effect of the system size. The number of oscillations increases with $d$, recall Fig.~\ref{fig:currents_d_vs_x}, and with the electronic density in the wire, $\mu-U_o$~\cite{extrinsic}. Let us concentrate on the case $\Delta=6\kBT$, for which we label them $\xi_o^a$ and $\xi_o^b$ in Fig.~\ref{fig:heat_lazos}(a). Those with largest currents, $\xi_o^a$, reduce the open channel current to a 15\%. Out of these points, the system explores a continuous set of configurations where the heat is routed to both output terminals simultaneously (and in different proportion) before one of them is blocked. Furthermore, in the way between the two switch points with equal currents in L and R, $\xi_\L^a$ and $\xi_\R^b$, the system necessarily passes again by the $\xi_o^b$ points with smaller output current. 
This flaw is avoided in shorter devices (as for $d<1.5l_0$ in Fig.~\ref{fig:currents_d_vs_x}) exhibiting a single vanishing point for each current, or by tuning the electrostatic potential of the wire with a plunger gate so as to reduce $\mu-U_0$. 

Alternatively, one can simply separate the tip from the conductor before displacing it and coupling it again in the desired position, in which case the system operates as a thermal SPDT switch controlled by two parameters, $x$ and $\epsilon$.

\subsection{Floating grounds}
We compare the short circuit mode of operation discussed in the above sections with the voltage probe source (when $I_\H=0$) and all-thermal (when all $I_j=0$) ones in Figs.~\ref{fig:heat_lazos}(b) and~\ref{fig:heat_lazos}(c).
Both boundary conditions restrict the values of $J_{\rm L}$ and $J_{\rm R}$ that are accessible by the system, with currents being smaller in the switching configurations, $\xi_\L$ and $\xi_R$. The two currents are correlated along the diagonal $J_{\rm L}\sim J_{\rm R}$ when $I_{\rm H}=0$ (compromising the router operation, as the input current increases in the process of switching), and close to the corner of the third quadrant (so that $J_{\rm R}\propto 1/J_{\rm L}$) in the all-thermal configuration. Note that adding constrictions to the particle flow makes the behaviour closer to the configuration where the resonances are not split ($\Delta=0$). This is because the floating electrochemical potentials of the leads tend to align with the position of the resonances. In the all thermal case, for instance, $\mu_\L$ and $\mu_\H$ will get close to $\varepsilon_1$ when $J_\L\neq0$ and $J_\R\approx0$. 

\section{Switching via gating
%Dephasing model
}
\label{sec:dephasing}

Most of the nanoscale semiconductor technologies are based on the control of the electrical properties by means of gate voltages, think e.g., on the field effect transistor, the quantum point contact or single-electron transistors~\cite{ihn_semiconductor_2009}. In this section we present a model of a SPDT switch based on a three-terminal junction that can be tuned by plunger gates. 
%%%%%%%%%%%%%%%%%%%%%%%%%%%%%%
\begin{figure}[t]
\includegraphics[width=\linewidth]{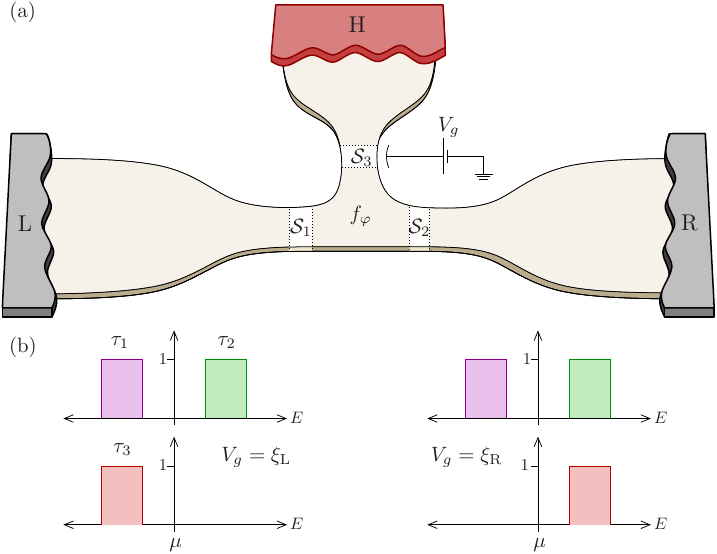}
\caption{\label{fig:schdephasing}
(a) Sketch of a mesoscopic region connected to three terminals, $\L$, $\R$ and $\H$, via scattering centers with scattering matrices ${\cal S}_\alpha$ acting as energy filters. Phase randomization in the central region results in a distribution $f_\varphi(E)$.
(b) Tuning a gate voltage $V_g$ applied to the filter connected to the heat source selects the output terminal for the injected electrons by matching the transmission function of the corresponding output filter.
}
\end{figure}
%%%%%%%%%%%%%%%%%%%%%%%%%%%%%%

The system is sketched in Fig.~\ref{fig:schdephasing}: three scattering centers, with scattering matrices ${\cal S}_\alpha$, $\alpha=1,2,3$, acting as energy filters for electrons injected from terminals L, R and H, respectively, are connected by a junction where we assume that electrons lose their phase coherence, stressing the difference with the interference-based mechanism discussed in Sec.~\ref{sec:coherent}~\cite{interfvsgate}. 
The idea is simple: if scatterers 1 and 2 filter particles flowing into the output terminals at non-coinciding energies, a gate voltage applied to the input scatterer, 3, can filter the energy of the injected particles from H such that they selectively flow either into terminal L or into terminal R, see Fig.~\ref{fig:schdephasing}(b). In this case, we write $\varepsilon_3=\varepsilon_3^0-eV_g$, where $\varepsilon_3^0$ is the bare resonance energy of the scatterer and $\xi=V_g$, the tuned parameter, is the voltage applied to a plunger gate close to it. If the two output filters are specularly situated across the electrochemical potential, ${\cal S}_1(E-\mu)={\cal S}_2(\mu-E)$, the output heat currents at the switch states $\xi_\L$ and $\xi_\R$ will be equal by construction. This can be achieved with additional plunger gates which we are ignoring here for simplicity. 

The connecting three terminal junction is subject to strong dephasing, such that there is no internal interference in the conductor. This is the case for instance of a cavity where electrons randomize their kinetic phase but do not suffer any inelastic scattering process~\cite{dejong_semiclassical_1996}. Classical junctions have also been considered~\cite{horvat_railway_2012}.
While recent works have proposed a measurement-based approach for treating dephasing with the proper fluctuation properties~\cite{sanchezFernan_monitored_2026},  it will suffice for our purposes here to use a quasielastic probe model accounting for the average currents~\cite{dejong_semiclassical_1996,Buttiker1991}.
The fictitious probe, labeled by $\varphi$, is treated as a fourth terminal that is connected to the three barriers and is assumed to absorb no current at all energies: ${\cal I}_\varphi(E)=0$. This translates to a boundary condition for the spectral current
\beq
{\cal I}_\varphi(E)=\frac{2}{h}\sum_j {\cal T}_{\varphi j}(E)[f_\varphi(E)-f_j(E)]=0,
\eeq
which results in a non-equilibrium distribution~\cite{vanLangen_quantum_1997}
\beq
f_\varphi(E)=\frac{\sum_j {\cal T}_{\varphi j}(E)f_j(E)}{\sum_j {\cal T}_{\varphi j}(E)}
\eeq
at the junction. 
With this, the spectral currents are written as:
\beq
{\cal I}_j(E)=\frac{2}{h}\sum_{j'}\tilde{\cal T}_{jj'}(E)[f_j(E)-f_{j'}(E)]
\eeq
in terms of the {\it effective} transmission probabilities 
\beq
\tilde{\cal T}_{jj'}(E)={\cal T}_{j\varphi}(E){\cal T}_{\varphi j'}(E)/\sum_k{\cal T}_{\varphi k}(E).
\eeq
The interpretation is clear: only terminals with a filter aligned with $\varepsilon_j$ will contribute to the currents $I_j$ and $J_j$. Note that the randomization of the kinetic phase imposes a finite backscattering probability: even for energies $E$ for which all scattering centers are transparent, ${\cal T}_{j\varphi}(E)={\cal T}_{\varphi j'}(E)=1$, the transmission probability is $\tilde{\cal T}_{jj'}(E)=1/2$.

\subsection{Resonant tunneling barriers}
\label{sec:restun}
The simplest such configuration is based on resonant tunneling barriers: only electrons with energies around the resonance are transmitted between the dephasing junction and terminal $j_\alpha$ with a probability ${\cal T}_{j_\alpha\varphi}(E)=|\tau_\alpha|^2$ given by Eq.~\eqref{eq:resonance}. We use here the labelling $j_1=\L$, $j_2=\R$ and $j_3=\H$. The gate voltages control the resonance energies, $\varepsilon_\alpha$.
If the splitting $\Delta \equiv\varepsilon_2-\varepsilon_1\neq0$, and assuming that $\Delta\gg\Gamma,\kBT$, matching the energy $\varepsilon_3$ with either of the two other resonances will result in current injected from H flowing into only one of the output reservoirs, hence acting as a thermal SPDT. The other output terminal will only exchange particles with the junction. 

%%%%%%%%%%%%%%%%%%%%%%%%%%%%%%
\begin{figure}[t]
\includegraphics[width=.9\linewidth]{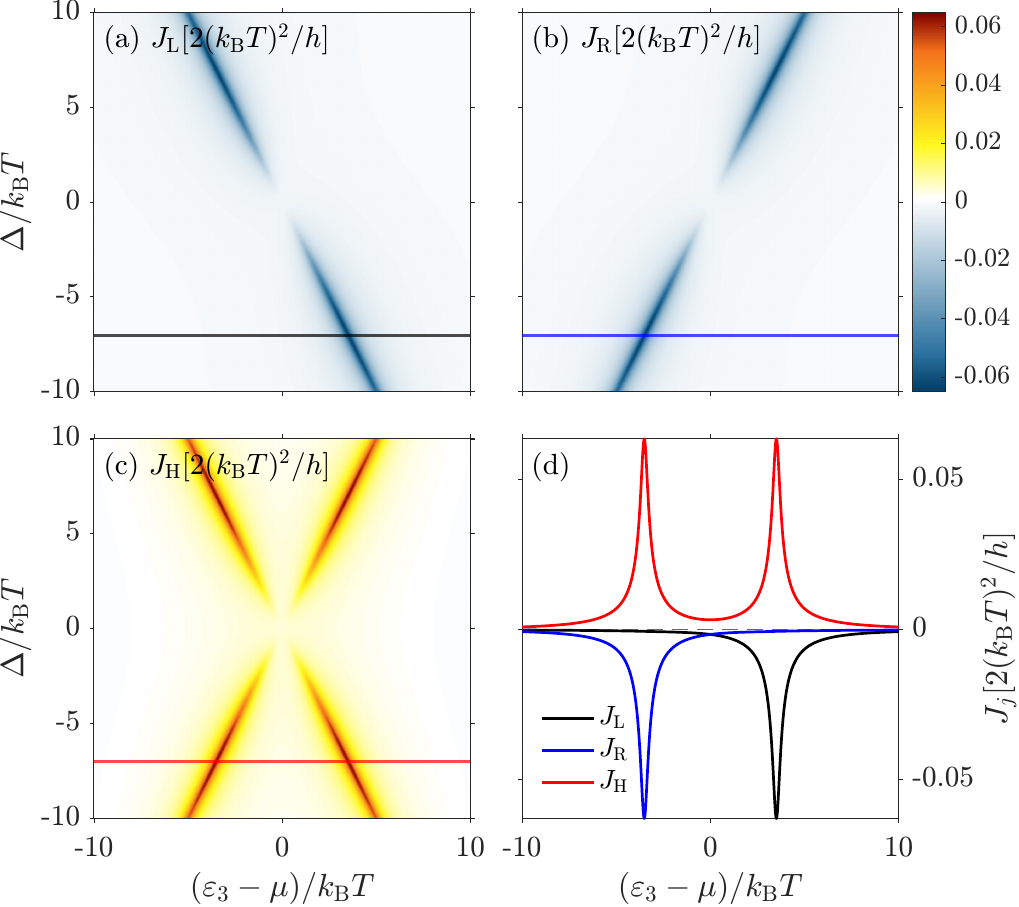}
\caption{\label{fig:IJdeph} (a)-(c) Heat currents as functions of the position of the resonance coupled to the hot terminal, $\varepsilon_3$, and of the splitting between the other two resonances, $\Delta$, with $\varepsilon_2=\mu+\Delta/2$ and $\varepsilon_1=\mu-\Delta/2$. (d) Cuts of the different currents for $\Delta=-7\kBT$. Parameters: $\Gamma=0.1\kBT$, $T_\H=2T$ and $\mu=0$.
}
\end{figure}
%%%%%%%%%%%%%%%%%%%%%%%%%%%%%%

The operation of the system as a thermal SPDT switch is shown in Fig.~\ref{fig:IJdeph}. We plot the heat currents as the resonance of the hot terminal barrier and the  splitting $\Delta$ are tuned around the electrochemical potential ($\varepsilon_1+\varepsilon_2=2\mu$). As expected, currents in terminals $\L$ and $\R$ are finite only when $\varepsilon_{1}\approx\varepsilon_3$ (configuration $\xi_\L$) or $\varepsilon_{2}\approx\varepsilon_3$ (configuration $\xi_\R$), with transport through the other terminal being suppressed. Since the two output barriers are chosen to have the same broadening, the maxima of the two output heat currents are equal for the same splitting $\Delta$, $J_\L(\varepsilon_3,\Delta)=J_\R(-\varepsilon_3,\Delta)$, see Figs.~\ref{fig:IJdeph}(a)-\ref{fig:IJdeph}(d). 

The performance of the switch is in this case limited by the overlap of the two resonance tails, as can be appreciated in Fig.~\ref{fig:IJdeph}(d) for a moderate splitting (not much larger than the resonance broadening). Increasing the splitting further the overlap is strongly suppressed. The optimal switching configuration is when $\Gamma<\Delta$ (so the overlap output resonances is suppressed) and $\Delta\sim \pm 3.5\kBT_\H$, where the heat spectral current $(\varepsilon_3-\mu)[f_\H(\varepsilon_3)-f_o(\varepsilon_3)]$ finds its maximum.
We note in passing that for this very symmetric configuration, terminal $\H$ injects only heat ($I_\H=0$, $J_\H\neq0$) when $\varepsilon_\H=\mu$, see Fig.~\ref{fig:IJdeph}(d) (the corresponding particle currents are shown in appendix~\ref{app:particleSPDT}): heat is carried into terminal R by electrons, and by holes into terminal L, as long as $\Delta\sim\Gamma>0$. Therefore at this point, the injected heat is converted in a finite particle current between terminals $\L$ and $\R$, in a dephasing version of a quantum thermocouple, for which one typically considers the conductor hotspot to be a voltage probe~\cite{jordan:2013,balduque_quantum_2026}.

%%%%%%%%%%%%%%%%%%%%%%%%%%%%%%
\begin{figure}[t]
\includegraphics[width=\linewidth]{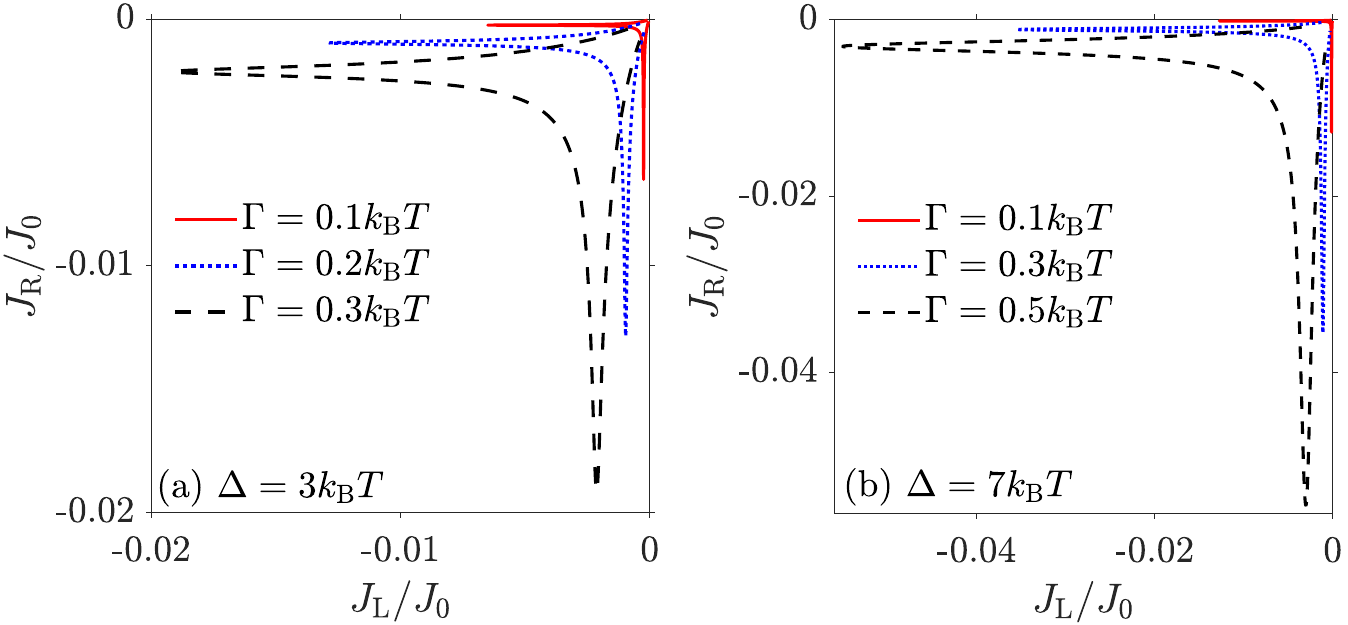}
\caption{\label{fig:ILRJLRdeph}
Heat currents as the position of the resonance coupled to the hot terminal, $\varepsilon_3$, is tuned, for different values of the broadening $\Gamma$ and for (a) $\Delta=3\kBT$ and (b) the maximal current splitting, $\Delta=7\kBT$. Other parameters are as in Fig.~\ref{fig:IJdeph}. 
%\textcolor{red}{Así está bien, pero mejor si las juntas y pones (a) y (b).}
}
\end{figure}
%%%%%%%%%%%%%%%%%%%%%%%%%%%%%%
The effect of the broadening is further explored in the parametric plots of Fig.~\ref{fig:ILRJLRdeph}, showing the heat currents when $\varepsilon_3$ is tuned for two different values of $\Delta$. As the broadening increases, the output currents increase. However, the degree of suppression of the other output current gets compromised due to the overlap of the tails of the resonances. This is particularly evident for small values of $\Delta$, so resonances are close to the electrochemical potential: not only currents currents are smaller, there is a non-negligible overlap of the output resonances harming the filtering effect with the increase of the broadening, $\Gamma$, see Fig.~\ref{fig:ILRJLRdeph}(a). The situation with higher splitting $\Delta\approx3.5\kBT_\H$ is more favorable to extract heat (resonances capture the maxima of the heat spectral current) and tolerates larger values of $\Gamma$ (and therefore even larger currents) without harming the switching off of one of them. For the particle currents, the thermoelectric peaks are slightly asymmetric, with tails being suppressed for energies closer to $\mu$, see appendix~\ref{app:particleSPDT}.

\subsection{Box-car shaped filters}
%%%%%%%%%%%%%%%%%%%%%%%%%%%%%%
\begin{figure}[t]
\includegraphics[width=.9\linewidth]{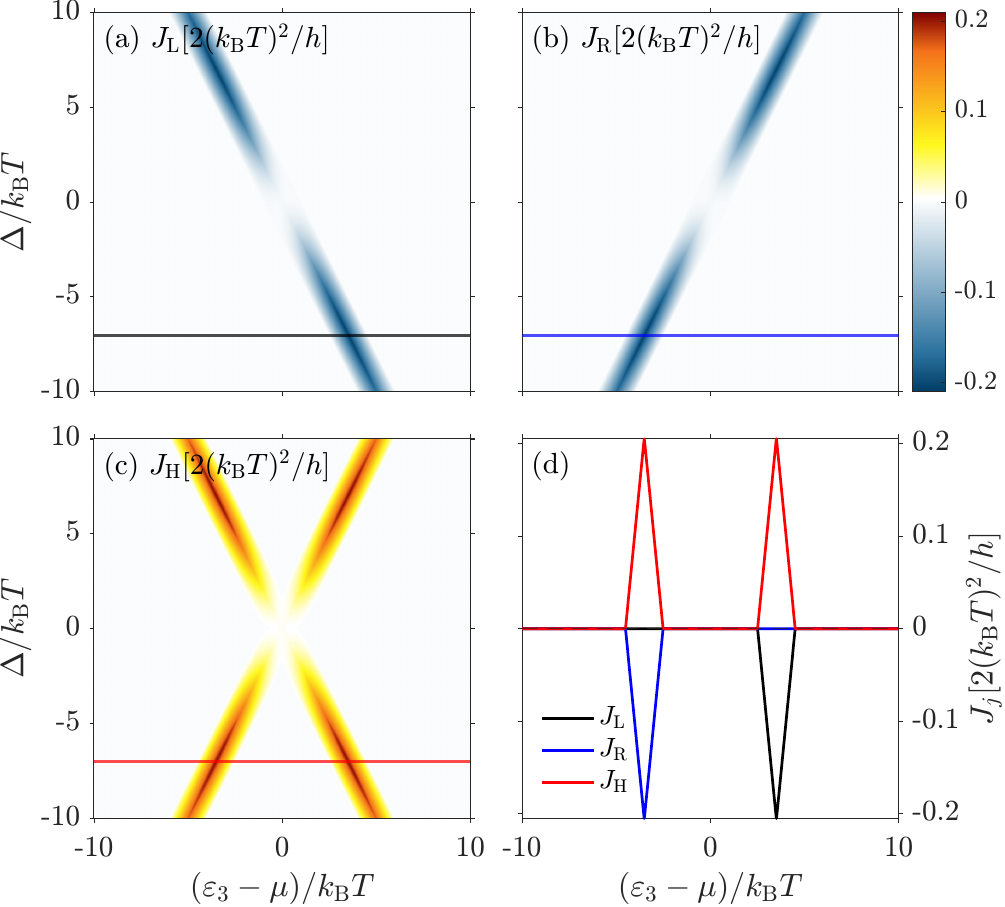}
\caption{\label{fig:IJboxcar} (a)-(c) Heat currents as functions of the central position of the box-car transmission coupled to the hot terminal, $\varepsilon_3$, and of the splitting between the other two box-cars, $\Delta$, with $\varepsilon_2=\mu+\Delta/2$ and $\varepsilon_1=\mu-\Delta/2$. (d) Cuts of the different currents for $\Delta=-7\kBT$. Parameters: $w=\kBT$, $T_\H=2T$ and $\mu=0$.
}
\end{figure}
%%%%%%%%%%%%%%%%%%%%%%%%%%%%%%

To improve the SPDT switch effect it is therefore beneficial to have scatterers with a wider filter window (allowing for larger currents) and sharper edges (so the overlap between the two output filters is suppressed). This is the case of box-car-shaped transmissions, which have been found to give highly efficient thermoelectric power~\cite{hershfield:2013,whitney_most_2014,balduque_coherent_2024}.
We explore this configuration by considering sharp transmissions ${\cal T}_{j_\alpha\varphi}(E)=|\tau_\alpha^{\rm bc}(E)|^2$, with
\beq
|\tau_\alpha^{\rm bc}(E)|^2=\Theta(E-\varepsilon_\alpha+w_\alpha/2)\Theta(\varepsilon_\alpha+w_\alpha/2-E),
\eeq
where $\Theta(E)$ is the Heaviside step function. The boxcar function of width $w_\alpha$ is centered at $\varepsilon_\alpha$. The currents are limited by the smallest width of the filters. For simplicity we here assume all filters to have equal width, $w_\alpha=w$, $\forall\alpha$. 
Then, if the transmissions $|\tau_1^{\rm bc}(E)|^2$ and $|\tau_2^{\rm bc}(E)|^2$ do not overlap, tuning the filter in lead H to coincide with one or the other opens an energy window of width $w$ for electrons to flow from H to one of the output terminals, $o$, with a probability $\tilde{\cal T}_{oH}=1/2$, while $\tilde{\cal T}_{o'H}=0$ for the other terminal. These are the situations sketched in Fig.~\ref{fig:schdephasing}(b).

The resulting currents are plotted in Fig.~\ref{fig:IJboxcar} as functions of the splitting of the conductor filters, $\Delta=\varepsilon_2-\varepsilon_1$, with $\varepsilon_1+\varepsilon_2=2\mu$, and of the energy of the input filter, $\varepsilon_3$. As soon as $\Delta>w$, the output filters do not overlap. Hence, shifting the input filter to $\varepsilon_3=\varepsilon_1$ or to $\varepsilon_3=\varepsilon_2$ brings all the heat injected from H to respectively flow into terminal L or R, see Figs.~\ref{fig:IJboxcar}(a)-\ref{fig:IJboxcar}(d). 

\begin{figure}[t]
\includegraphics[width=\linewidth]{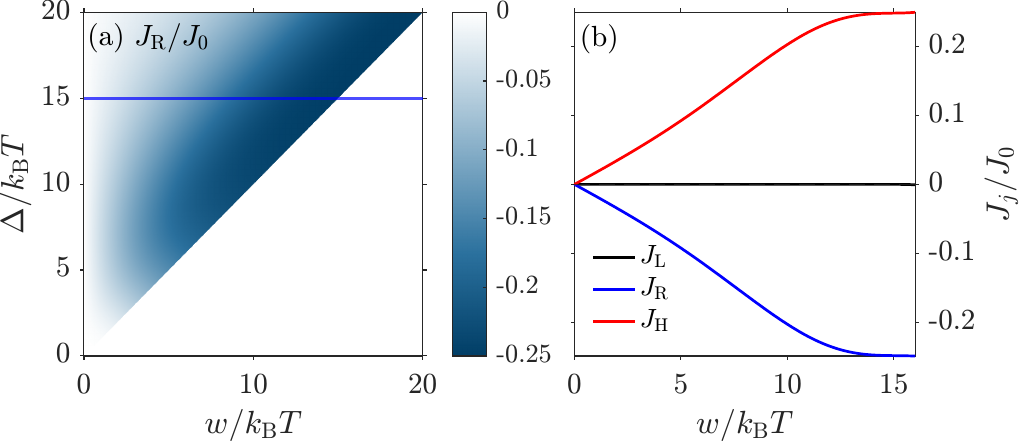}
\caption{\label{fig:IJdephboxw}
(a) Output heat current at R, $J_\R$, as a function of the boxcar width $w$ and the splitting $\Delta$, with $\varepsilon_2=\mu+\Delta/2$ and $\varepsilon_1=\mu-\Delta/2$, and $\varepsilon_3=\varepsilon_2$. Only the region with $w\leq\Delta$, where the thermal SPDT switching is perfect, is plotted. (b) Cuts of the different currents for $\Delta=15\kBT$. Parameters: $T_\H=2T$, $\mu=0$.
}
\end{figure}

The maximal response occurs when the input and output filters are perfectly aligned, so we expect it to be only limited by the filter width. To have equal output currents, filters 1 and 2 need to be antisymmetric: $|\tau_1^{\rm bc}(E-\mu)|^2=|\tau_2^{\rm bc}(\mu-E)|^2$. The largest amount of heat will hence be when the width of the box-car is large $w\gg\kBT$, so one of them acts effectively as a positive $E-\mu$ pass filter (like a high-pass filter at $E=\mu$), the other one being a negative $E-\mu$ filter (a low-pass filter at $\mu$). Taking this and the fact that electrons in the filtered windows are reflected with a probability 1/2 into account, we find that the current will be bounded by $J_0/4$. For an $N_o$-channel output lead, hence
\beq
-J_o\leq N_o\pi^2 k_{\rm B}^2(T_\H^2-T^2)/12h. 
\eeq
The onset of this behaviour occurs when the filter is able to capture the region of maximal spectral current $(E-\mu){\cal I}_\H$, which occurs for $\varepsilon_3-\mu=w/2\gtrsim3.5\kBT_\H$ as long as $T_\H\gtrsim T$~\cite{boxcarmax}. 
This is confirmed by Fig.~\ref{fig:IJdephboxw}, showing the output current in terminal R as a function of the splitting $\Delta$ as the condition for $\xi_\R$ is maintained, i.e., $\varepsilon_3=\varepsilon_2$, while $w$ is also tuned. For any given $\Delta$, the maximal output current is obtained at $\Delta=w$, as expected, with the current saturating for large $\Delta$ at $J_0/4$: further increasing $\Delta$ opens for transport a region with exponentially vanishing spectral heat current $(E-\mu){\cal I}_\H$.

%\begin{figure}[t]
%\includegraphics[width=0.8\linewidth]{boxSPDT_IJvseHD.pdf}
%\caption{\label{fig:IJdephbox}
%(a)-(c) Particle and (e-)-(g) heat currents as functions of the position of the boxcar coupled to the hot terminal, $\varepsilon_H$ and of the splitting between the other two resonances, $\Delta$, with $\varepsilon_R=\mu+\Delta/2$ and $\varepsilon_L=\mu-\Delta/2$. Cuts of the different currents for $\Delta=2.5\kBT$ are plotted in (d) and (h). Parameters: $\Gamma=0.1\kBT$, $T_H=2T$, $\mu=0$, $w=5.5\kBT$, $\gamma=0.2\kBT$.
%}
%\end{figure}

\section{Conclusions}
\label{sec:conclusions}
We have proposed three-terminal junctions as thermal (and thermoelectric) routers and SPDT switches in an mesoscopic electronic conductor. One of the terminals injects heat, and the other two control the transfer of particles and heat via energy filters, e.g. resonant tunneling barriers. Two mechanisms are explored: 
(i) a fully coherent one based on the global interference pattern arising from the relative position of the particle injection with respect to the filters, and (ii) a dephased one based on filtering at the source terminal as well. 

The first case (i) can be implemented with the coherent injection of a scanning tip between the two conductor barriers. The interference pattern can be tuned via the position of the tip thanks to the interplay of Fabry-P\'erot- and Fano-like contributions. This way the ratio of the two output currents can be controlled continuously. For particular tip positions, the transmission to the output terminals can be selectively suppressed. We compare the performance for three different configurations depending on the conditions of the reservoir electrochemical potentials: grounded terminals permit larger heat currents but work only at specific points; open-circuit conditions give less conducting but more robust switches. To avoid the need of movable components, possible alternatives include a fixed tri-junction tuning the kinetic phases in branches l and r with applied gate voltages or using the Aharonov-Bohm phase of annular junctions~\cite{adrian}.

The second case (ii) is simpler to implement and requires no movable parts. Junctions based on resonant tunneling are very efficient thermal SPDT switches but are limited by the filter broadening: narrow filters transfer small currents, but wider resonances overlap. The optimal configuration requires wide and sharp spectral filters, e.g., with the shape of a box-car, for which the ovelap is suppressed and currents are large. In this case we obtain an upper bound of the output current given by 1/4 of the heat current allowed in an open two-terminal conductor with the same temperatures.

%%%%%%%%%%%%%%%%%%%%%%%%%%%%%%%%%%%%%%%
\acknowledgements
We acknowledge funding from the Spanish Ministerio de Ciencia e Innovaci\'on via grants  No. PID2022-142911NB-I00 and No. PID2024-157821NB-I00, and through the ``Mar\'{i}a de Maeztu'' Programme for Units of Excellence in R{\&}D CEX2023-001316-M.

\appendix

\section{Thermoelectric SPDT switch}
\label{app:particleSPDT}

%\sout{The switching behavior is more clearly seen in the cuts at fixed $d$ plotted in Figs.~\ref{fig:MechanicalPhE_valve_currents}(d) and \ref{fig:MechanicalPhE_valve_currents}(h): by changing the position of the tip we switch from a situation with finite currents at R and zero output in L to the opposite one. }

The thermoelectric switch has a salient property with respect to the thermal version discussed in the main text: the flow of particles between terminals at different temperatures depends on whether it is dominated by particles or by holes~\cite{benenti_thermodynamic_2011}. This means that, even if heat always flows from the input to the output terminals (remember Clausius), particles may flow in different directions through the output terminals (e.g., from H to L when $\xi=\xi_\L$ and from R to H when $\xi=\xi_\R$). 
%We will call a {\it strict} switch the one where terminal H is the source of both particles and heat in both switch positions, $\xi_\L$ and $\xi_\R$.

\subsection{Interferometer case}

\begin{figure}[t]
    \centering
    \includegraphics[width=\linewidth]{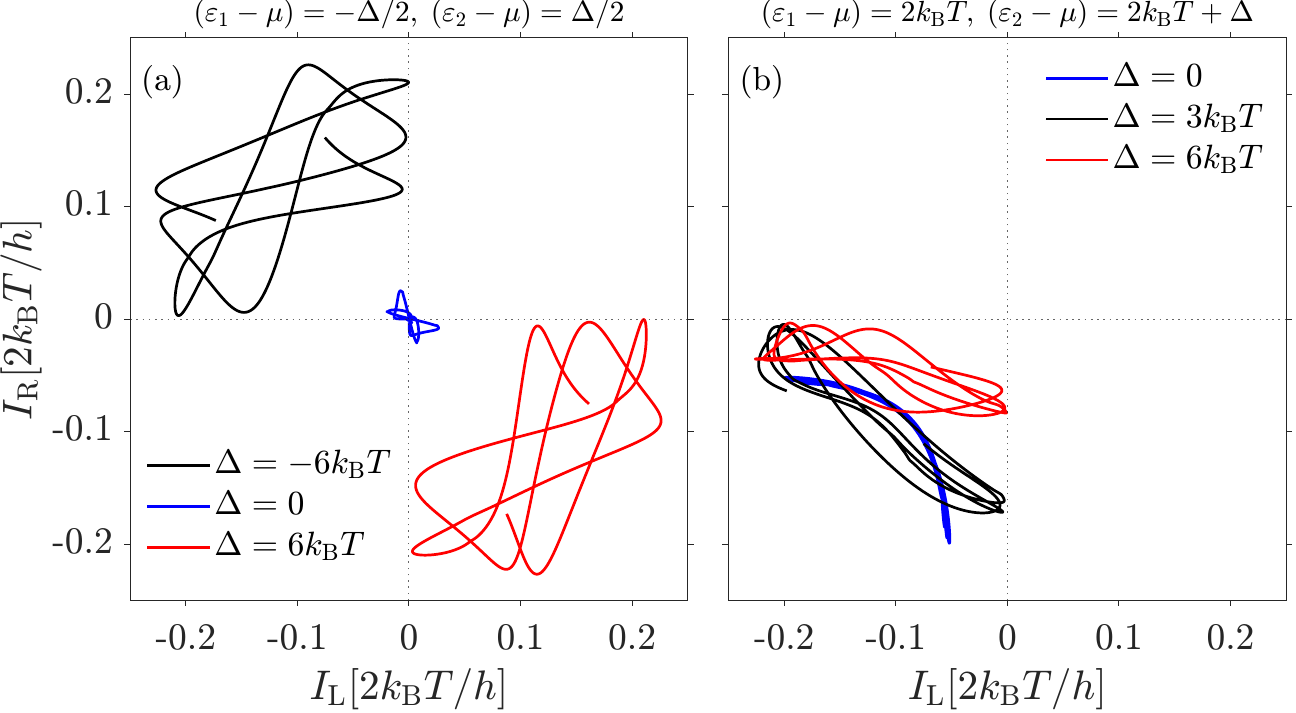}
   \caption{\label{fig:particle_lazos} Particle currents as the position of the tip position, $x$ is tuned, with fixed $d/l_0=3.5$ and different $\Delta$, for (a) the same parameters of Fig.~\ref{fig:heat_lazos}(a), and (b) $\varepsilon_1-\mu=2\kBT$, $\varepsilon_2-\mu=2\kBT+\Delta$.  }
\end{figure}

In this appendix we discuss the particle currents for the coherent setup discussed in Sec.~\ref{sec:coherent}, in particular how the nonlocal thermoelectric response of the device determines their sign in the short-circuit configuration with $\mu_j=\mu$. For this we compare the cases in which the energies of the two resonant barriers are either antisymmetric with respect to the ground electrochemical potential (the same configuration discussed in Sec.~\ref{sec:coherent}) or both over the electrochemical potential.

Let us first consider the case where they are asymmetrically situated with respect to the electrochemical potential, e.g., with $\varepsilon_2>\mu>\varepsilon_1$. Then, when ${\cal T}_{LH}\approx0$, transport between H and R is dominated by hot electrons and hence $I_H>0$. On the contrary, when ${\cal T}_{RH}\approx0$, transport between H and L is dominated by holes and hence $I_H<0$. This can be appreciated in Fig.~\ref{fig:particle_lazos}(a), which reproduces the {\it arabesque}-like lines of the heat currents in Fig.~\ref{fig:heat_lazos}(a), with the switched-on currents having different signs. Reversing $\Delta$, the sign of both currents change. At $\Delta=0$, when $\varepsilon_1=\varepsilon_2=\mu$, an oscillating thermoelectric response appears due to the internal interference~\cite{extrinsic}.

In the case where both barrier resonances, $\varepsilon_1$ and $\varepsilon_2$ are over the Fermi energy, the two currents are dominated by electrons flowing from H to the output terminals, see Fig.~\ref{fig:particle_lazos}(b). Then, all particle and heat currents flow from H to the output terminals.
Again in the symmetric case with $\Delta=0$, the interference due to multiple reflection processes between barriers is able to make the two currents asymmetric. However, the overlap of the two resonances avoids the suppression of either one, see blue curve in Fig.~\ref{fig:particle_lazos}(b). 
When both barrier resonances are negative, particles flow in the opposite directions (not shown).

%In this appendix we plot the particle and heat currents for the coherent setup discussed in Sec.~\ref{sec:coherent} in the case where the energies of the two resonant barriers are both over the electrochemical potential, $\varepsilon_\alpha>\mu$, see Figs.~\ref{fig:MechanicalPhE_valve_currentsd} and \ref{fig:MechanicalPhE_valve_currentsDE.

\subsection{Incoherently connected resonances}

%%%%%%%%%%%%%%%%%%%%%%%%%%%%%%
\begin{figure}[t]
\includegraphics[width=.9\linewidth]{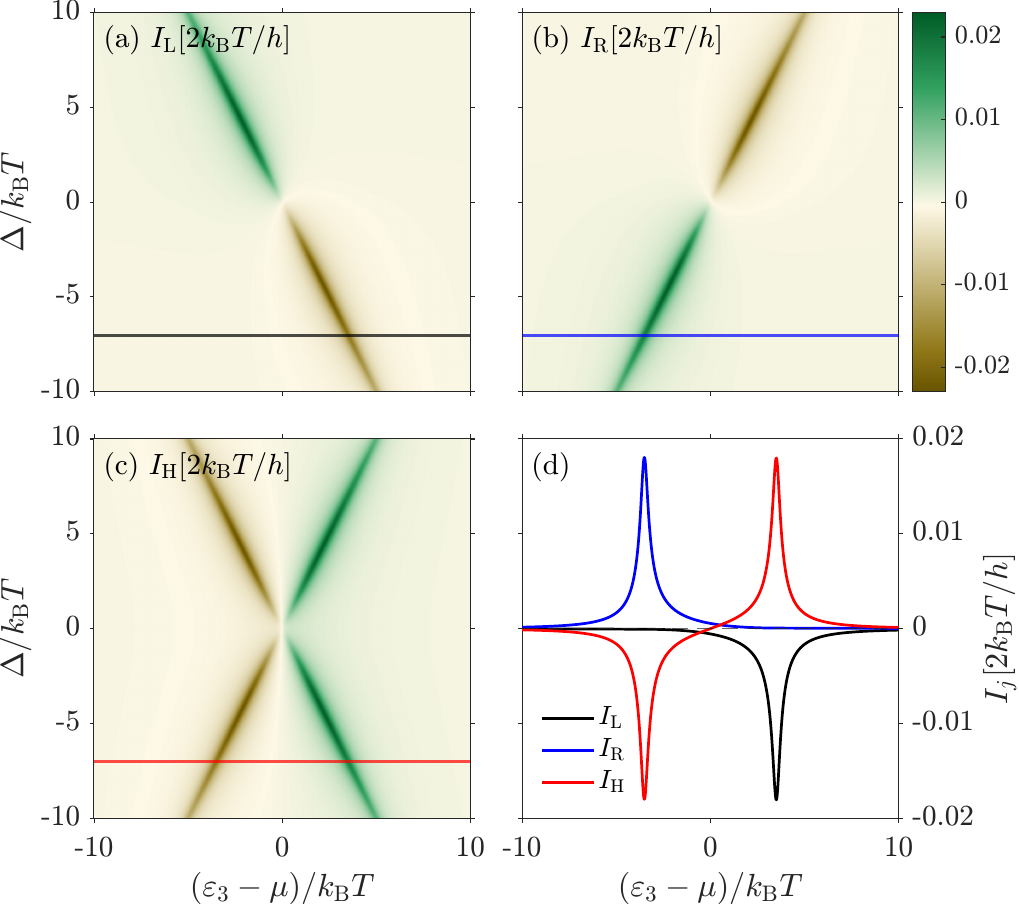}
\caption{\label{fig:Ideph}
(a)-(c) Particle currents as functions of the position of the resonance coupled to the hot terminal, $\varepsilon_3$ and of the splitting between the other two resonances, $\Delta$, with $\varepsilon_2=\mu+\Delta/2$ and $\varepsilon_1=\mu-\Delta/2$. (d) Cuts of the different currents for $\Delta=-7\kBT$. Parameters: $\Gamma=0.1\kBT$, $T_\H=2T$ and $\mu=0$.
}
\end{figure}
%%%%%%%%%%%%%%%%%%%%%%%%%%%%%%

We now consider the particle currents in the gate-tunable system formed by resonant-tunneling junctions discussed in Sec.~\ref{sec:restun}.
Choosing the output resonance energies to be symmetric around the electrochemical potential, $\varepsilon_2-\mu=\mu-\varepsilon_1$, the particle currents that flow into $\L$ and $\R$ for $\xi=\xi_\L$ and $\xi=\xi_\R$ for the same $\Delta$ have opposite signs, $I_\L(\varepsilon_\H,\Delta)=-I_\R(-\varepsilon_\H,\Delta)$, see Figs.~\ref{fig:Ideph}(a)-\ref{fig:Ideph}(d). Different from heat, the particle current $I_\H$ vanishes in this configuration when $\varepsilon_\H=\mu$ because junctions 1 and 2 generate currents of equal magnitude but opposite signs in terminals L and R~\cite{balduque_quantum_2026}.

%%%%%%%%%%%%%%%%%%%%%%%%%%%%%%
\begin{figure}[t]
\includegraphics[width=0.6\linewidth]{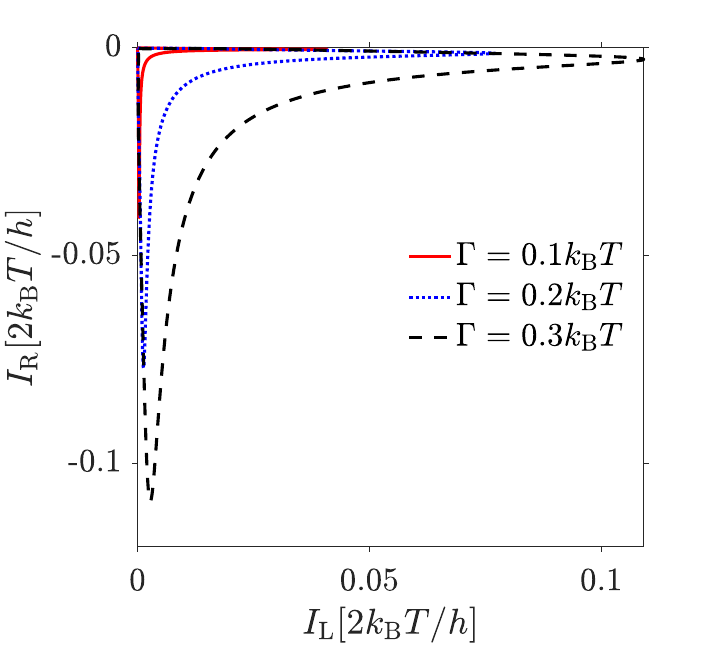}
\caption{\label{fig:ILRdeph}
Particle currents as the position of the resonance coupled to the hot terminal, $\varepsilon_3$ is tuned, with $\Delta=3\kBT$ and for different values of $\Gamma$, with other parameters as in Fig.\ref{fig:IJdeph}. 
}
\end{figure}
%%%%%%%%%%%%%%%%%%%%%%%%%%%%%%

A parametric plot of the output currents as the switching parameter is tuned is shown in Fig.~\ref{fig:ILRdeph}. The device behaviour resembles the thermal case, plotted in Fig.~\ref{fig:ILRJLRdeph}(a). This is expected in the narrow-resonance limit ($\Gamma\ll\kBT$) where particle and heat flows are proportional to each other.

%%%%%%%%%%%%%%%%%%%
%%%%%%%%%%%%%%%%%%%
%%%%%%%%%%%%%%%%%%%
%%%%%%%%%%%%%%%%%%%
%%%%%%%%%%%%%%%%%%%
%%%%%%%%%%%%%%%%%%%
%%%%%%%%%%%%%%%%%%%
%%%%%%%%%%%%%%%%%%%
%%%%%%%%%%%%%%%%%%%
%%%%%%%%%%%%%%%%%%%

%%%%%%%%%%%%%%%%%%%%%%%%%%%%%%%%%%%%%%%%%%%%%%%%%%%%%%%%%%%%%%%%%%%%%%%%%%%%%%

\bibliography{biblio.bib}

\begin{thebibliography}{71}
\bibitem [{int()}]{interfvsgate}%
  \BibitemOpen
  \href@noop {} {}\bibinfo {note} {Note however that the interference pattern
  discussed in Sec.~\ref{sec:coherent} can also be tuned by gating the local
  electrostatic potential landscape of the wire.}\BibitemShut {Stop}%
\end{thebibliography}
%%%%%%%%%%%%%%%%%%%%%%%%%%%%%%%%%%%%%%%

%\begin{figure*}[t]
%    \centering
%    \includegraphics[width=0.7\linewidth]{current_d_vs_x_top.pdf}
   %\caption{\label{fig:MechanicalPhE_valve_currentsd} (a)-(c) Particle and (e)-(f) heat currents when the temperature of reservoir H is increased by $\Delta T/T=1$, as a function of the separation between scatterers $d$ and the tip position $x$. Panels (d) and (e) show cuts of the previous for fixed distance $d=2l_0$, indicated with horizontal lines. Parameters: $\varepsilon_1-\mu=2\kBT$, $\varepsilon_2-\mu=6\kBT$, $\epsilon=0.5$, $\Gamma=0.5\kBT$ and $\mu=20\kBT$. }
%\end{figure*}

%\begin{figure*}[t]
%    \centering
%    \includegraphics[width=0.7\linewidth]{current_DeltaE_top_vs_x.pdf}
   %\caption{\label{fig:MechanicalPhE_valve_currentsDE} (a)-(c) Particle and (e)-(f) heat currents when the temperature of reservoir H is increased by $\Delta T/T=1$, as a function of the tip position $x$, and \textcolor{red}{\sout{$\varepsilon_2-\mu=\Delta$} $\Delta=\varepsilon_2-\varepsilon_1$ (?)}. Panels (d) and (e) show cuts of the previous for fixed energy gain indicated with horizontal lines. Parameters: $d/l_0=3.5\varepsilon_1=22\kBT$, $\epsilon=0.5$, $\Gamma=0.5\kBT$ and $\mu=20\kBT$. }
%\end{figure*}

\end{document}